\documentclass[showpacs,prb,twocolumn]{revtex4}
\usepackage{epsfig}
\begin{document}
\title{Kinetic glass behavior in a diffusive model}
\author{D.~\v Capeta and D.K.~Sunko\thanks{email: dks\@phy.hr}}
\affiliation{
Department of Physics,\\Faculty of Science,\\
University of Zagreb,\\
Bijeni\v cka cesta 32,\\
HR-10000 Zagreb, Croatia.}

\begin{abstract}

Three properties of the Edwards-Anderson model with mobile bonds are
investigated which are characteristic of kinetic glasses. First is two-time
relaxation in aged systems, where a significant difference is observed between
spin and bond autocorrelation functions. The spin subsystem does not show
two-time behavior, and the relaxation is stretched exponential. The bond
subsystem shows two-time behavior, with the first relaxation nearly
exponential and the second similar to the spin one. Second is the
two-temperature behavior, which can be tuned by bond dilution through the full
range reported in the literature. Third is the rigid-to-floppy transition,
identified as a function of bond dilution. Simple Glauber Monte Carlo
evolution without extraneous constraints reproduces the behavior of classical
kinetic simulations, with the bond (spin) degree of freedom corresponding to
configurational (orientational) disorder.

\end{abstract}


\maketitle

Kinetic glasses are long-lived nonequilibrium systems of considerable
technological importance. The basic insight into their microscopic origin was
provided by Kauzmann:\cite{Kauzmann48} there is a slowdown in configurational
rearrangement, caused by obstruction of kinetic motion. Despite a consensus in
this view and advances in theoretically enlightened phenomenology of known
glass formers,\cite{Angell95,Debenedetti01} there is at present no unifying
view of vitrification. This is in remarkable contrast with its practical
simplicity: put honey into the freezer or let egg whites dry, and a kinetic
glass appears without effort.

Out of equilibrium, subdominant microscopic correlations multiply, and
finding the one responsible for macroscopic structural arrest is even more
difficult than doing so for equilibrium transitions. Simulations of abstract
model systems have developed into an important tool to extract generic
behavior in such a situation, although of course they do not substitute for
direct physical insight into real
cases.\cite{Thorpe00,Hehlen02,Simon06,Boolchand01,Boolchand01-1} While many
models have been found which exhibit some kind of slowed-down response, their
relationship with one another, and with physical reality, is still the subject
of continuous investigation.

Molecular dynamics (MD) simulations of two-component Lennard-Jones fluids are
intuitively closest to real systems.\cite{Kob95} They have been
compared with the predictions of mode-coupling
theory,\cite{Gotze75,Bosse78,Flenner05} a high-temperature microscopic
approach, which depends on the explicit introduction of a three-particle
scattering term.

More abstract, and schematic, are constrained kinetics
simulations.\cite{Toninelli06} These are lattice gas models in which updates
are not dictated by pure coupling to the thermostat, but are supplemented with
special rules, standing in for higher-order correlations, which otherwise
cannot be included in a random walk.

A separate class of models, sometimes used to understand vitrification, are
models of spin glasses.\cite{Cugliandolo02} It requires some abstraction, or
leap of faith, to identify the former microscopic physical spin with a
hypothetical ``mesoscopic'' order parameter, representative of
the slowed-down configuratonal rearrangement. Nevertheless, particular
quenched systems have remarkable parallels with other models of kinetic
vitrification, the reasons for which have lately become better
understood.\cite{Crisanti00,Jund01,Rao03,Moore06}

Monte Carlo (MC) simulations play an important role in many of these
investigations. In the time domain, they simplify joint probabilities as
$p(x_1,t_1;x_2,t_2;x_3,t_3)=p(x_1,t_1;x_2,t_2)$ for times $t_1>t_2>t_3$, since
the underlying random walk has no memory. Here $x_i$ is the system
configuration and the times are measured with a resolution $\Delta\tau$
shorter than the thermalization time, but long enough for all correlations,
retaining information of the initial conditions, to die out, except the
two-time ones. Such a random walker is a minimal model of fluctuations in a
concrete thermostat. Since physical three-particle correlations are generally
three-time correlations, in ensemble language one may say MC simulations take
two-particle physical correlations as input and build higher correlations
statistically as output, on scales coarser than $\Delta\tau$.

In this work we address the issue as to how much needs to be said physically
to obtain vitrification statistically. From this point of view, all of the
approaches above suffer from some surfeit. Full Newtonian evolution does all
the work physically, with no lower limit on $\Delta\tau$ in principle. Mode
coupling has a three-particle kernel as input. Constrained kinetics has
memory, since an update can overrule the thermostat. Conversely, to enforce
the rules by the energy balance requires multibody forces, stronger than
two-body. Random quenched disorder cannot be the input, since the
comparatively well-organized vitreous disorder~\cite{Phillips79,Phillips81} is
a consequence of evolution and by a Hamiltonian which has an ordered ground
state.

Our main result is that less needs to be said than suspected so far. We apply
a natural annealment dynamics to the Edwards-Anderson model for spin
glasses.\cite{Edwards75} While the equilibrium state of the model was
described~\cite{Thorpe76,Thomsen86} long ago, it turns out the approach to
equilibrium is meaningful for kinetic glasses. When nearest-neighbor bond
diffusion is added to spin flips, vitreous delay appears in the ordinary
random walk. The difference with most other theoretical approaches is that
there are two different degrees of freedom in interaction, so the physical
correlation responsible for vitrification appears to be a second-order
off-diagonal term (mixed in conjugate fields). In contrast to other spin-glass
models, it turns out that the bond, not spin, degree of freedom corresponds to
the configurational one. Bond movement gives the model an unexpected
``off-lattice'' character.


We study the two-dimensional short-range Edwards-Anderson
model,\cite{Edwards75} as before,~\cite{Lazic01,Capeta05} at $B=0$:
\begin{equation}
H=-\sum_{\left<i,j\right>}J_{ij}S_iS_j-M\cdot B,
\label{model}
\end{equation}
where $S_i=\pm 1$ and we shall use both a Gaussian distribution of $J_{ij}$
with half-width $J$ and a bimodal $\pm J$ distribution. We investigate a
two-dimensional lattice of size $500\times 500$ with Glauber dynamics. Between
each two spin trials (flips) is a bond trial: a positive and negative bond
impinging on the same site, and chosen at random, are allowed to exchange
places, subject to the same criterion as the spin trials. The concentration
$p_{AF}$ of antiferromagnetic (AF) bonds is taken to be 50\%.

Bond updates quickly anneal the sample, and after a transient of $\sim$100
updates per site it enters a long-lived state with glasslike
dynamics,\cite{Lazic01} and a second, much longer relaxation
time.\footnote{Fig.~6 of Ref.~\onlinecite{Lazic01} refers to spins only, and
begins at the quench, unlike Fig.~1 here, taken long after the quench.} The
``glassiness'' of the metastable spin state is not topological: the underlying
bond distribution has low frustration at any instant in time, so much
so that it can be mapped onto a disordered ferromagnet,\cite{Hartmann03} with
a finite transition temperature.\cite{Merz02} Bond diffusion neverthless
prevents the spins from settling into any given ordered state, leading to a
decay of correlations even below the hidden phase transition.

When bonds move, the model theoretically evolves towards an annealed
equilibrium.\cite{Thorpe76,Thomsen86} However, as long as bond updates are
kept local, their diffusion is not very efficient in finding the optimal
configurations, despite efficient annealment in energy. We do not establish
directly that the configurations are suboptimal, but when bond hops of
arbitrary range are allowed, the configurations obtained are quite different
than those found here, having a strong tendency to dropletlike phase
separation.\cite{Lazic99} Although equilibration is delayed from the point of
view of correlations, evolution passes only through a small subclass of
energetically favorable states, though bond updates imply that not only the
equilibrium spin manifold is being sampled. This is similar to the situation
in real kinetic glasses, while in spin glasses the average configuration
energy is much higher.\cite{Lazic01} We find that isotropic spatial
correlation functions fall off rapidly within a few lattice spacings, in
accordance with recent thinking.\cite{Tarzia06}


\begin{figure}
\center{\epsfig{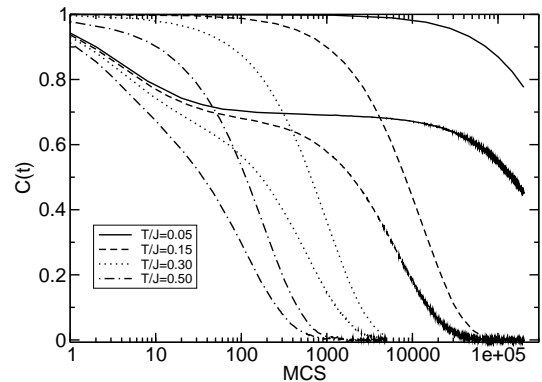}}
\caption{Spin and bond autocorrelation functions in an aged sample. In each
pair of curves, the bond autocorrelation is the one with the plateau at
intermediate times. The full, dashed, dotted, and dot-dashed curves correspond
to temperatures $T/J=0.05$, 0.15, 0.30, and 0.50, respectively. Each Monte
Carlo step (MCS) consists of a spin and a bond update per site.}
\label{figautocorr}
\end{figure}
In Fig.~\ref{figautocorr}, we show the autocorrelation functions
of two copies of an aged sample, for spins given by
\begin{equation}
C(t)={1\over V}\sum_{i=1}^V\left<S_i(t_0)S_i(t_0+t)\right>,
\label{ct}
\end{equation}
and similarly for bonds, for the Gaussian distribution. ``Aged'' means $t_0$
long enough for no transients to survive after quenching, so the system
is in the long-lived metastable state. The spin and bond autocorrelation
functions are different. The spin autocorrelation decays as a stretched
exponential $\exp\left[-(t/\tau)^\beta\right]$, which is the usual behavior
in disordered spin models. The bond autocorrelation shows
obvious two-time behavior, with a plateau separating the fast, closer to
exponential decay ($\beta\approx 0.9$), from a slow, stretched exponential
component ($\beta\approx 0.8$). The plateau is characteristic of
configurational relaxation in kinetic glasses and observed in classical
simulations of binary systems.\cite{Kob95} The fast component ``shakes out''
those configurations which may be relaxed locally, while the slow component
refers to ``locked-in'' ones, whose relaxation is impeded by intermediate-range
correlations. We recall~\cite{Kauzmann48} that the orientational degree of
freedom is generally not affected by the glass transition, being slowed down
already above $T_g$. As the temperature is lowered, the spin and bond
responses decouple ever more, which is visible in the figure as an extended
plateau in the bond response. This is also analogous to real systems, where
there are more translations relative to rotations the deeper one goes into
the glassy regime.\cite{Debenedetti01} If spin updates are discontinued, the
bond response loses the second relaxation component, and the plateau
then extends to the longest times investigated. The reason is that fixed
annealed spins create barriers of satisfied bonds, which mobile bonds find
difficult to cross, so the system is broken into uncommunicating regions. When
spin updates are allowed, these regions themselves evolve slowly, accounting
for the second relaxation time. Rare spin updates at low temperature
correspond to activated hops in constrained kinetics, absent from
mode coupling, which similarly retains the plateau \emph{ad infinitum}.

\begin{figure}
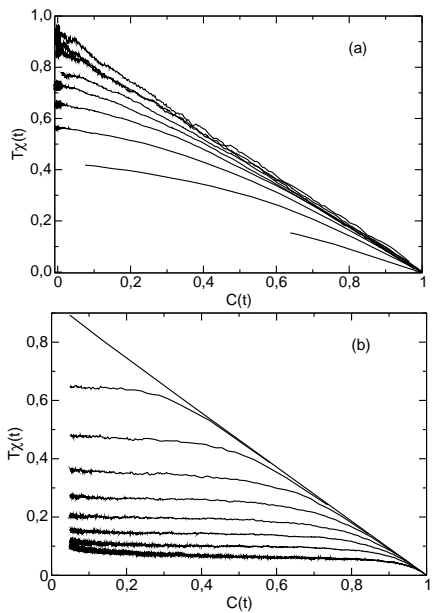

\center{\epsfig{file=fig_fdtg,height=40mm}
\epsfig{file=fig_fdtj0,height=40mm}}
\caption{Violations of the fluctuation-dissipation theorem for an (a) Gaussian
and (b) bimodal distribution. (a) Bottom to top:
$T/J=0.05,0.01,0.015,0.2,0.25,0.3,0.4,0.5,0.6$, respectively. (b) Top to
bottom: $p_0=$ 100\%, 90\%, 80\%, 70\%, 60\%, 40\%, 30\%, 20\%, 10\%,
respectively, at $T/J=0.2$.}
\label{figtwotemp}
\end{figure}
Next, we investigate how the interplay of the two degrees of freedom affects
departures from the fluctuation-dissipation theorem (FDT), expected in the
same context.\cite{Cugliandolo02} Here we plot $C(t)$ for spins [Eq.
(\ref{ct})] against the spin susceptibility normalized by temperature:
\begin{equation}
T\chi(t)={T\over V}
\lim_{\Delta B\to 0}\left.{\Delta M\over\Delta B}\right|_{t_0+t},
\end{equation}
for $\Delta B(t_0+t)$ a step function at $t=0$. Although such violations are
well established for models with quenched disorder, it is not clear that
annealing caused by bond diffusion will not quickly lead to equilibration. In
fact, it does not, and for the same Gaussian bond distribution as above, we
obtain in Fig.~\ref{figtwotemp} the ``many-temperature'' violation curve, known
from previous investigations of spin models and observed in some
experiments.\cite{Herisson02} Since the model is only coupled to a simple
thermostat, the violation is evidently generated internally.

Even more interesting is the case of the $\pm J$ bond distribution. We
introduce bond dilution through a fixed proportion $p_0$ of bonds with $J=0$.
For $p_0=1$, an ordinary paramagnet, the FDT is evidently observed. As the
dilution is reduced there are again violations, but instead of many
temperatures, the ``two-temperature'' shape appears, familiar from
MD simulations of two-sphere models.\cite{Berthier02}
However, in the $\pm J$ case there is no plateau in the bond response,
indicating that a fine energy scale is needed for the creation of inherent
structures.

\begin{figure}
\center{\epsfig{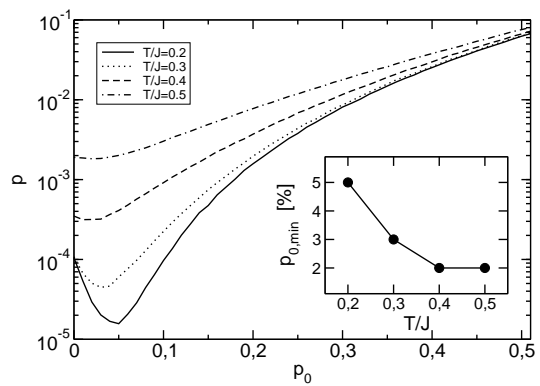}}
\caption{Rigid-to-floppy transition. The fraction of indifferent spins is
shown as a function of bond dilution. The minimum corresponds to the
least-strained system. The inset shows the minimum as a function of
temperature.}
\label{figbool}
\end{figure}
The rigid-to-floppy transition has emerged as an important paradigm in a wide
range of glass formers.\cite{Thorpe00,Boolchand01} The basic
idea~\cite{Phillips79,Phillips81} is that a glassy system can be either
overdetermined (rigid) or underdetermined (floppy), with some ideal
``unstrained-but-stiff'' configuration in the middle. Figure~\ref{figbool}
shows this model''s version of the rigid-to-floppy transition. It gives the
fraction $p$ of spins in a zero local field (``indifferent'') as a function of
bond dilution $p_0$. As $p_0\to 1$ (paramagnet), all spins become indifferent:
$p(1)=1$ is the extreme floppy limit. However, the curves $p(p_0)$ also show a
sharp minimum at a finite $p_0$. This is not difficult to understand: an
indifferent spin in an undiluted system occurs wherever two positive and two
negative bonds impinge on the same site. It can be oriented by putting a
single zero bond in place of one of them. Hence the number of indifferent
spins initially decreases with dilution. The minimum corresponds to the
least-strained state in the sense above: to the left, $p$ increases because
there are too many constraints and to the right, because there are too few.
The trend of the minimum with temperature is also reasonable; one expects that
the lower the temperature of the quench, the more zero bonds are needed to
effectively unstrain the system.


Experimental evidence suggests that vitrification is a two-stage process. The
first stage is dominated by quantum mechanics---i.e.\ chemical bonding---which
determines what is ``locally favorable.'' Once the first stage has given rise
to entities at the scale of $\sim$10--100~nm, their further evolution may be
imagined in classical or geometrical terms. Most investigations study this
second stage of vitrification, including the present work. The paradigm of a
macroscopic internally generated potential, borrowed from phase transitions,
is evidently sensible, since something is certainly precluding window glass
from free flow. However, the universally observed liquid structure
factor~\cite{Kelton06} means this potential varies in a way irreconcilable
with a divergent length scale. Thus the notion of an internally generated
field is more general than that of a phase transition. In addition,
macroscopic approaches,\cite{Kivelson95,Chamon05} based on dynamic scaling
ideas, indicate that the time scales of the internal field are finite as well.

Given that classical simulations do not at present correspond to any
well-defined physical objects---indeed the ``configurational degree of
freedom'' is as intuitive a concept today as when its importance was first
noted~\cite{Kauzmann48}---it is reasonable to ask whether their success in
reproducing particular observed aspects of glassy dynamics comes from the side
of the thermostat, rather than being a higher-order effect of precise
mechanical evolution. From this point of view, ordinary MC simulations are a
fundamental tool. They build up high-order correlations without microscopic
trajectories, enabling one to study how close to equilibrium glasses really
are.

The present work obtains a vitreous response with a minimal disturbance of the
annealed spin equilibrium and the minimal physical input to date. All that
seems required is dynamical evolution of two different degrees of freedom with
a local interaction. Hence vitrification in the model is due to a second-order
off-diagonal correlation, rather than a higher-order diagonal one, typical of
models with one kind of degree of freedom. It raises the possibility that the
same could be true in real systems, such as large molecules, the
interdependence of translation and rotation (both kinetic) in a log-jam coming
readily to mind. The idea of configurational rearrangement agrees with the
interpretation of off-diagonal correlations as conditional probabilities: a
particle cannot move \emph{unless} another one does. However, the definite
identification of the correlation involved depends on finding a natural
conjugate field, a nontrivial task, even in a simple model such as ours.
Reproducing MD by pure MC simulations~\cite{Berthier06} is very similar to the
present result; we have removed from it one more physical prop, the realistic
potential, so it only seems important that there be two kinds of little balls.
This fits well with binary (and ternary) mixtures being easier to vitrify than
pure substances.~\cite{Boolchand01,Boolchand01-1}

Another issue is that the model has a hidden ferromagnetic
transition.\cite{Hartmann03,Merz02} It is analogous to the avoided
crystallization in vitreous liquids. The bond autocorrelation implies that
discontinuing spin updates triggers a true transition, as is known to happen
in mode coupling.

To conclude, we have given a concrete example of vitrification in a simple
model, where it is by default due to a second-order off-diagonal correlation.
The appearance of vitreous slowdown in the bond response of the
Edwards-Anderson model when bond diffusion is introduced points to its kinetic
origin. We believe this type of correlation may be responsible for
vitrification in at least some real glasses, where it would provide the
minimal physical content of the configurational degree of freedom.


Comments by J.~C.~Phillips are gratefully acknowledged. This work was
supported by the Croatian Government under Project No.~$0119256$.


\begin{thebibliography}{35}
\expandafter\ifx\csname natexlab\endcsname\relax\def\natexlab#1{#1}\fi
\expandafter\ifx\csname bibnamefont\endcsname\relax
  \def\bibnamefont#1{#1}\fi
\expandafter\ifx\csname bibfnamefont\endcsname\relax
  \def\bibfnamefont#1{#1}\fi
\expandafter\ifx\csname citenamefont\endcsname\relax
  \def\citenamefont#1{#1}\fi
\expandafter\ifx\csname url\endcsname\relax
  \def\url#1{\texttt{#1}}\fi
\expandafter\ifx\csname urlprefix\endcsname\relax\def\urlprefix{URL }\fi
\providecommand{\bibinfo}[2]{#2}
\providecommand{\eprint}[2][]{\url{#2}}

\bibitem[{\citenamefont{Kauzmann}(1948)}]{Kauzmann48}
\bibinfo{author}{\bibfnamefont{W.}~\bibnamefont{Kauzmann}},
  \bibinfo{journal}{Chem. Rev.} \textbf{\bibinfo{volume}{43}},
  \bibinfo{pages}{219} (\bibinfo{year}{1948}).

\bibitem[{\citenamefont{Angell}(1995)}]{Angell95}
\bibinfo{author}{\bibfnamefont{C.}~\bibnamefont{Angell}},
  \bibinfo{journal}{Science} \textbf{\bibinfo{volume}{267}},
  \bibinfo{pages}{1924} (\bibinfo{year}{1995}).

\bibitem[{\citenamefont{Debenedetti and Stillinger}(2001)}]{Debenedetti01}
\bibinfo{author}{\bibfnamefont{P.~G.} \bibnamefont{Debenedetti}}
  \bibnamefont{and} \bibinfo{author}{\bibfnamefont{F.~H.}
  \bibnamefont{Stillinger}}, \bibinfo{journal}{Nature}
  \textbf{\bibinfo{volume}{410}}, \bibinfo{pages}{259} (\bibinfo{year}{2001}).

\bibitem[{\citenamefont{Thorpe et~al.}(2000)\citenamefont{Thorpe, Jacobs,
  Chubynsky, and Phillips}}]{Thorpe00}
\bibinfo{author}{\bibfnamefont{M.~F.} \bibnamefont{Thorpe}}
  \bibnamefont{et~al.}, \bibinfo{journal}{J. Non-Cryst. Solids}
  \textbf{\bibinfo{volume}{266-269}}, \bibinfo{pages}{859}
  (\bibinfo{year}{2000}).

\bibitem[{\citenamefont{Hehlen et~al.}(2002)\citenamefont{Hehlen, Courtens,
  Yamanaka, and Inoue}}]{Hehlen02}
\bibinfo{author}{\bibfnamefont{B.}~\bibnamefont{Hehlen}}
  \bibnamefont{et~al.}, \bibinfo{journal}{J. Non-Cryst. Solids}
  \textbf{\bibinfo{volume}{307}}, \bibinfo{pages}{87} (\bibinfo{year}{2002}).

\bibitem[{\citenamefont{Simon et~al.}(2006)\citenamefont{Simon, Hehlen,
  Courtens, Longueteau, and Vacher}}]{Simon06}
\bibinfo{author}{\bibfnamefont{G.}~\bibnamefont{Simon}}
  \bibnamefont{et~al.}, \bibinfo{journal}{Phys. Rev. Lett.}
  \textbf{\bibinfo{volume}{96}}, \bibinfo{pages}{105502}
  (\bibinfo{year}{2006}).

\bibitem[{\citenamefont{Boolchand
  et~al.}(2001{\natexlab{a}})\citenamefont{Boolchand, Georgiev, and
  Goodman}}]{Boolchand01}
\bibinfo{author}{\bibfnamefont{P.}~\bibnamefont{Boolchand}}
  \bibnamefont{et~al.}, \bibinfo{journal}{J. Optoel. Adv. Mat.}
  \textbf{\bibinfo{volume}{3}}, \bibinfo{pages}{703}
  (\bibinfo{year}{2001}{\natexlab{a}}).

\bibitem[{\citenamefont{Boolchand
  et~al.}(2001{\natexlab{b}})\citenamefont{Boolchand, Feng, and
  Bresser}}]{Boolchand01-1}
\bibinfo{author}{\bibfnamefont{P.}~\bibnamefont{Boolchand}}
  \bibnamefont{et~al.},
  \bibinfo{journal}{J. Non-Cryst. Solids} \textbf{\bibinfo{volume}{293-295}},
  \bibinfo{pages}{348} (\bibinfo{year}{2001}{\natexlab{b}}).

\bibitem[{\citenamefont{Kob and Andersen}(1995)}]{Kob95}
\bibinfo{author}{\bibfnamefont{W.}~\bibnamefont{Kob}} \bibnamefont{and}
  \bibinfo{author}{\bibfnamefont{H.~C.} \bibnamefont{Andersen}},
  \bibinfo{journal}{Phys. Rev. E} \textbf{\bibinfo{volume}{51}},
  \bibinfo{pages}{4626} (\bibinfo{year}{1995}).

\bibitem[{\citenamefont{{G\"otze} and {L\"ucke}}(1975)}]{Gotze75}
\bibinfo{author}{\bibfnamefont{W.}~\bibnamefont{{G\"otze}}} \bibnamefont{and}
  \bibinfo{author}{\bibfnamefont{M.}~\bibnamefont{{L\"ucke}}},
  \bibinfo{journal}{Phys. Rev. A} \textbf{\bibinfo{volume}{11}},
  \bibinfo{pages}{2173} (\bibinfo{year}{1975}).

\bibitem[{\citenamefont{Bosse et~al.}(1978)\citenamefont{Bosse, {G\"otze}, and
  {L\"ucke}}}]{Bosse78}
\bibinfo{author}{\bibfnamefont{J.}~\bibnamefont{Bosse}}
  \bibnamefont{et~al.}, \bibinfo{journal}{Phys. Rev. A}
  \textbf{\bibinfo{volume}{17}}, \bibinfo{pages}{434} (\bibinfo{year}{1978}).

\bibitem[{\citenamefont{Flenner and Szamel}(2005)}]{Flenner05}
\bibinfo{author}{\bibfnamefont{E.}~\bibnamefont{Flenner}} \bibnamefont{and}
  \bibinfo{author}{\bibfnamefont{G.}~\bibnamefont{Szamel}},
  \bibinfo{journal}{Phys. Rev. E} \textbf{\bibinfo{volume}{72}},
  \bibinfo{pages}{031508} (\bibinfo{year}{2005}).

\bibitem[{\citenamefont{Toninelli et~al.}(2006)\citenamefont{Toninelli, Biroli,
  and Fisher}}]{Toninelli06}
\bibinfo{author}{\bibfnamefont{C.}~\bibnamefont{Toninelli}}
  \bibnamefont{et~al.}, \bibinfo{journal}{Phys. Rev. Lett.}
  \textbf{\bibinfo{volume}{96}}, \bibinfo{pages}{035702}
  (\bibinfo{year}{2006}).

\bibitem[{\citenamefont{Cugliandolo}(2003)}]{Cugliandolo02}
\bibinfo{author}{\bibfnamefont{L.~F.} \bibnamefont{Cugliandolo}}, in
  \emph{\bibinfo{booktitle}{Slow Relaxations and Nonequilibrium Dynamics in
  Condensed Matter}}, edited by \bibinfo{editor}{\bibfnamefont{J.-L.}
  \bibnamefont{Barrat}}
  \bibnamefont{et~al.}
  (\bibinfo{publisher}{Springer}, \bibinfo{year}{2003}),
  vol.~\bibinfo{volume}{77} of \emph{\bibinfo{series}{Les Houches - Ecole d'Ete
  de Physique Theorique}}.

\bibitem[{\citenamefont{Crisanti and Ritort}(2000)}]{Crisanti00}
\bibinfo{author}{\bibfnamefont{A.}~\bibnamefont{Crisanti}} \bibnamefont{and}
  \bibinfo{author}{\bibfnamefont{F.}~\bibnamefont{Ritort}},
  \bibinfo{journal}{Physica A} \textbf{\bibinfo{volume}{280}},
  \bibinfo{pages}{155} (\bibinfo{year}{2000}).

\bibitem[{\citenamefont{Jund et~al.}(2001)\citenamefont{Jund, Jullien, and
  Campbell}}]{Jund01}
\bibinfo{author}{\bibfnamefont{P.}~\bibnamefont{Jund}}
  \bibnamefont{et~al.}, \bibinfo{journal}{Phys. Rev. E}
  \textbf{\bibinfo{volume}{63}}, \bibinfo{pages}{036131}
  (\bibinfo{year}{2001}).

\bibitem[{\citenamefont{Rao et~al.}(2003)\citenamefont{Rao, Crisanti, and
  Ritort}}]{Rao03}
\bibinfo{author}{\bibfnamefont{F.}~\bibnamefont{Rao}}
  \bibnamefont{et~al.}, \bibinfo{journal}{Europhys. Lett.}
  \textbf{\bibinfo{volume}{62}}, \bibinfo{pages}{869} (\bibinfo{year}{2003}).

\bibitem[{\citenamefont{Moore and Yeo}(2006)}]{Moore06}
\bibinfo{author}{\bibfnamefont{M.~A.} \bibnamefont{Moore}} \bibnamefont{and}
  \bibinfo{author}{\bibfnamefont{J.}~\bibnamefont{Yeo}},
  \bibinfo{journal}{Phys. Rev. Lett.} \textbf{\bibinfo{volume}{96}},
  \bibinfo{pages}{095701} (\bibinfo{year}{2006}).

\bibitem[{\citenamefont{Phillips}(1979)}]{Phillips79}
\bibinfo{author}{\bibfnamefont{J.~C.} \bibnamefont{Phillips}},
  \bibinfo{journal}{J. Non-Cryst. Solids} \textbf{\bibinfo{volume}{34}},
  \bibinfo{pages}{153} (\bibinfo{year}{1979}).

\bibitem[{\citenamefont{Phillips}(1981)}]{Phillips81}
\bibinfo{author}{\bibfnamefont{J.~C.} \bibnamefont{Phillips}},
  \bibinfo{journal}{J. Non-Cryst. Solids} \textbf{\bibinfo{volume}{43}},
  \bibinfo{pages}{37} (\bibinfo{year}{1981}).

\bibitem[{\citenamefont{Edwards and Anderson}(1975)}]{Edwards75}
\bibinfo{author}{\bibfnamefont{S.~F.} \bibnamefont{Edwards}} \bibnamefont{and}
  \bibinfo{author}{\bibfnamefont{P.~W.} \bibnamefont{Anderson}},
  \bibinfo{journal}{J. Phys. F.} \textbf{\bibinfo{volume}{5}},
  \bibinfo{pages}{965} (\bibinfo{year}{1975}).

\bibitem[{\citenamefont{Thorpe and Beeman}(1976)}]{Thorpe76}
\bibinfo{author}{\bibfnamefont{M.}~\bibnamefont{Thorpe}} \bibnamefont{and}
  \bibinfo{author}{\bibfnamefont{D.}~\bibnamefont{Beeman}},
  \bibinfo{journal}{Phys. Rev. B} \textbf{\bibinfo{volume}{14}},
  \bibinfo{pages}{188} (\bibinfo{year}{1976}).

\bibitem[{\citenamefont{Thomsen et~al.}(1986)\citenamefont{Thomsen, Thorpe,
  Choy, Sherrington, and Sommers}}]{Thomsen86}
\bibinfo{author}{\bibfnamefont{M.}~\bibnamefont{Thomsen}}
  \bibnamefont{et~al.}, \bibinfo{journal}{Phys. Rev. B}
  \textbf{\bibinfo{volume}{33}}, \bibinfo{pages}{1931} (\bibinfo{year}{1986}).

\bibitem[{\citenamefont{Lazi\'{c} and Sunko}(2001)}]{Lazic01}
\bibinfo{author}{\bibfnamefont{P.}~\bibnamefont{Lazi\'{c}}} \bibnamefont{and}
  \bibinfo{author}{\bibfnamefont{D.~K.} \bibnamefont{Sunko}},
  \bibinfo{journal}{Eur. Phys. J. B} \textbf{\bibinfo{volume}{21}},
  \bibinfo{pages}{595} (\bibinfo{year}{2001}).

\bibitem[{\citenamefont{{\v Capeta} and Sunko}(2005)}]{Capeta05}
\bibinfo{author}{\bibfnamefont{D.}~\bibnamefont{{\v Capeta}}} \bibnamefont{and}
  \bibinfo{author}{\bibfnamefont{D.~K.} \bibnamefont{Sunko}},
  \bibinfo{journal}{J. Magn. Magn. Mater.} \textbf{\bibinfo{volume}{292}},
  \bibinfo{pages}{359} (\bibinfo{year}{2005}).

\bibitem[{\citenamefont{Hartmann}(2003)}]{Hartmann03}
\bibinfo{author}{\bibfnamefont{A.~K.} \bibnamefont{Hartmann}},
  \bibinfo{journal}{Phys. Rev. B} \textbf{\bibinfo{volume}{67}},
  \bibinfo{pages}{214404} (\bibinfo{year}{2003}).

\bibitem[{\citenamefont{Merz and Chalker}(2002)}]{Merz02}
\bibinfo{author}{\bibfnamefont{F.}~\bibnamefont{Merz}} \bibnamefont{and}
  \bibinfo{author}{\bibfnamefont{J.~T.} \bibnamefont{Chalker}},
  \bibinfo{journal}{Phys. Rev. B} \textbf{\bibinfo{volume}{65}},
  \bibinfo{pages}{054425} (\bibinfo{year}{2002}).

\bibitem[{\citenamefont{Lazi\'{c}}(1999.)}]{Lazic99}
\bibinfo{author}{\bibfnamefont{P.}~\bibnamefont{Lazi\'{c}}},
  \bibinfo{publisher}{Diploma thesis, Faculty of
  Science, University of Zagreb}, \bibinfo{year}{1999.}

\bibitem[{\citenamefont{Tarzia and Moore}(2006)}]{Tarzia06}
\bibinfo{author}{\bibfnamefont{M.}~\bibnamefont{Tarzia}} \bibnamefont{and}
  \bibinfo{author}{\bibfnamefont{M.~A.} \bibnamefont{Moore}}
  (\bibinfo{year}{2006}), \bibinfo{note}{cond-mat/0609113}.

\bibitem[{\citenamefont{Herisson and Ocio}(2002)}]{Herisson02}
\bibinfo{author}{\bibfnamefont{D.}~\bibnamefont{Herisson}} \bibnamefont{and}
  \bibinfo{author}{\bibfnamefont{M.}~\bibnamefont{Ocio}},
  \bibinfo{journal}{Phys. Rev. Lett.} \textbf{\bibinfo{volume}{88}},
  \bibinfo{pages}{257202} (\bibinfo{year}{2002}).

\bibitem[{\citenamefont{Berthier and Barrat}(2002)}]{Berthier02}
\bibinfo{author}{\bibfnamefont{L.}~\bibnamefont{Berthier}} \bibnamefont{and}
  \bibinfo{author}{\bibfnamefont{J.-L.} \bibnamefont{Barrat}},
  \bibinfo{journal}{J. Chem. Phys.} \textbf{\bibinfo{volume}{116}},
  \bibinfo{pages}{6228} (\bibinfo{year}{2002}).

\bibitem[{\citenamefont{Kelton}(2006)}]{Kelton06}
\bibinfo{author}{\bibfnamefont{K.~F.} \bibnamefont{Kelton}},
  \bibinfo{journal}{Intermetallica} \textbf{\bibinfo{volume}{14}},
  \bibinfo{pages}{966} (\bibinfo{year}{2006}).

\bibitem[{\citenamefont{Kivelson et~al.}(1995)\citenamefont{Kivelson, Kivelson,
  Zhao, Nussinov, and Tarjus}}]{Kivelson95}
\bibinfo{author}{\bibfnamefont{D.}~\bibnamefont{Kivelson}}
  \bibnamefont{et~al.}, \bibinfo{journal}{Physica~A}
  \textbf{\bibinfo{volume}{219}}, \bibinfo{pages}{27} (\bibinfo{year}{1995}).

\bibitem[{\citenamefont{Chamon and Cugliandolo}(2005)}]{Chamon05}
\bibinfo{author}{\bibfnamefont{C.}~\bibnamefont{Chamon}} \bibnamefont{and}
  \bibinfo{author}{\bibfnamefont{L.~F.} \bibnamefont{Cugliandolo}},
  \bibinfo{journal}{Pramana-J. Phys.} \textbf{\bibinfo{volume}{64}},
  \bibinfo{pages}{1075} (\bibinfo{year}{2005}).

\bibitem[{\citenamefont{Berthier and Kob}(2006)}]{Berthier06}
\bibinfo{author}{\bibfnamefont{L.}~\bibnamefont{Berthier}} \bibnamefont{and}
  \bibinfo{author}{\bibfnamefont{W.}~\bibnamefont{Kob}} (\bibinfo{year}{2006}),
  \bibinfo{note}{cond-mat/0610253}.

\end{thebibliography}
\end{document}